\documentclass[twocolumn,superscriptaddress,preprintnumbers,amsmath,amssymb,prl]{revtex4}
\UseRawInputEncoding
\usepackage{graphicx}

\begin{document}
\title{Casimir effect for magnetic media: Spatially nonlocal response
to the off-shell quantum fluctuations}

\author{
G.~L.~Klimchitskaya}
\affiliation{Central Astronomical Observatory at Pulkovo of the
Russian Academy of Sciences, Saint Petersburg,
196140, Russia}
\affiliation{Institute of Physics, Nanotechnology and
Telecommunications, Peter the Great Saint Petersburg
Polytechnic University, Saint Petersburg, 195251, Russia}

\author{
V.~M.~Mostepanenko}
\affiliation{Central Astronomical Observatory at Pulkovo of the
Russian Academy of Sciences, Saint Petersburg,
196140, Russia}
\affiliation{Institute of Physics, Nanotechnology and
Telecommunications, Peter the Great Saint Petersburg
Polytechnic University, Saint Petersburg, 195251, Russia}
\affiliation{Kazan Federal University, Kazan, 420008, Russia}

\begin{abstract}
We extend the Lifshitz theory of the Casimir force to the case
of two parallel magnetic metal plates possessing a spatially
nonlocal dielectric response. By solving Maxwell equations in
the configuration of an electromagnetic wave incident on the
boundary plane of a magnetic metal semispace, the exact surface
impedances are  expressed in terms of its magnetic permeability
and longitudinal and transverse dielectric functions. This allows
application of the Lifshitz theory with reflection coefficients
written via the surface impedances for calculation of the Casimir
pressure between magnetic metal (Ni) plates whose dielectric
responses are described by the alternative nonlocal response
functions introduced for the case of nonmagnetic media.
It is shown that at separations from 100 to 800~nm the Casimir
pressures computed using the alternative nonlocal and local plasma
response functions differ by less than 1\%. At separations of a
few micrometers, the predictions of these two approaches differ
between themselves and between that one obtained using the Drude
function by several tens of percent. We also compute the gradient
of the Casimir force between Ni-coated surfaces of a sphere and
a plate using the alternative nonlocal response functions and
find a very good agreement with the measurement data.
Implications of the obtained results determined by the off-shell
quantum fluctuations to a resolution of long-standing problems
in the Casimir physics are discussed.
\end{abstract}

\maketitle

\newcommand{\kb}{{k_{\bot}}}
\newcommand{\skb}{{k_{\bot}^2}}
\newcommand{\vk}{{\mbox{\boldmath$k$}}}
\newcommand{\rv}{{\mbox{\boldmath$r$}}}
\newcommand{\ve}{{\varepsilon}}
\newcommand{\Te}{{\varepsilon^{\rm Tr}}}
\newcommand{\Le}{{\varepsilon^{\,\rm L}}}
\newcommand{\Tle}{{\varepsilon_l^{\rm Tr}}}
\newcommand{\Lle}{{\varepsilon_l^{\,\rm L}}}
\newcommand{\zo}{{(z;\omega,k_x)}}
\newcommand{\oz}{{(\omega,k_x,k_z)}}
\newcommand{\xk}{{(i\xi_l,k_{\bot})}}
\newcommand{\VT}{{v^{\rm Tr}}}
\newcommand{\VL}{{v^{\,\rm L}}}
\section{Introduction}

An attractive force between two parallel uncharged ideal metal planes
in vacuum was predicted by H.~B.~G.~Casimir \cite{1} and is referred
to by his name. As an effect caused by the zero-point oscillations of
quantum fields, the Casimir force found a wide application in both
elementary particle physics and cosmology. Specifically, the Casimir
energy of quark and gluon fields contributes some part of the total
energy of hadrons in the bag model \cite{2,3}. The Casimir effect
provides a mechanism for the compactification of extra dimensions in
Kaluza-Klein field theories \cite{4}, affects the evolution of
cosmological models with nontrivial topology \cite{5,6}, and allows
to place strong constraints on non-Newtonian gravity and light
elementary particles \cite{7,8,9}. The Casimir force has also
become the topic of a large body of research in atomic and condensed
matter physics \cite{10,11,12,13,14,15,16}.

There are two main approaches to theory of the Casimir effect. The
first of them, which goes back to Casimir \cite{1}, is based on
quantum field theory. In order to find the Casimir energy in the
framework of this approach, one should consider the quantum field in
a restricted quantization volume, determine the energy eigenvalues,
sum them up, and apply the appropriate regularization and renormalization
procedures for obtaining the finite result \cite{1,17,18,19,20,21}.
The second approach, which is based on quantum statistical physics,
goes back to Lifshitz \cite{22,23}. This approach uses the concept of
a fluctuating field created by stochastic currents existing inside
the bodies bounding the quantization volume. According to the
fluctuation-dissipation theorem, the spectral distribution of
fluctuations is expressed via the imaginary part of a response
function of the boundary materials to quantum fluctuations which
permits to find an expression for the stress tensor and finally for
the Casimir interaction.

Both approaches lead to the Lifshitz formulas for the Casimir free
energy and force between two thick plates (semispaces) described by
the frequency-dependent dielectric permittivities as response
functions. In Ref. \cite{24} the Lifshitz formulas were generalized
to the case of magnetic media. There is, however, an important
difference between the two approaches. The quantum field theoretical
approach is the most rigorous when the boundary problem under
consideration has real eigenvalues. This is the case for the
ideal metal boundaries, in applications to the elementary particle
physics and cosmology, and also for some idealized dielectrics and
metals whose dielectric functions are constant or described by the
dissipationless plasma model, respectively. To derive the Lifshitz
formula for more realistic boundary bodies possessing dissipation,
the quantum field theoretical approach was combined with some
auxiliary electrodynamic problem \cite{25}. By contrast, the
statistical physics derivation results in the Lifshitz formula
solely for the dissipative media where the dielectric function
possesses a nonzero imaginary part leading to the complex
eigenvalues of the boundary problem. This is in rather poor agreement
with the fact that a substitution of real dielectric permittivity
of the plasma model in the Lifshitz formula results in a nonzero
Casimir force.

Repeated precise experiments on measuring the Casimir interaction
between metallic test bodies \cite{26,27,28,29,30,31,32,33,34,35,36,37,38}
revealed a puzzling problem. It turned out that theoretical predictions
of the Lifshitz theory are excluded by the measurement data if the
dielectric response of a metal at low frequencies is described by the
well-tested dissipative Drude function possessing a nonzero imaginary part,
as required by the statistical physics derivation of the Lifshitz formula.
The same experiments \cite{26,27,28,29,30,31,32,33,34,35,36,37,38} were
found to be in a very good agreement with calculations using the Lifshitz
formula if the low-frequency dielectric response of the boundary bodies
is described by the real plasma function which disregards dissipation and
should be inapplicable at low frequencies (at sufficiently high
frequencies, where the optical data of interacting bodies are available,
the response functions along the imaginary frequency axis
in both cases were found using the Kramers-Kronig
relations from the measured complex index of refraction
\cite{11,13,14,15,16}).

It is meaningful also that the Lifshitz theory using the Drude response
function violates the Nernst heat theorem for metals with perfect crystal
lattice which is a truly equilibrium system with a nondegenerate ground
state \cite{39,40,41,42} (an agreement is restored for only the crystal
lattices containing some fraction of impurities \cite{43,44,45}). The
Lifshitz theory using the plasma response function satisfies the Nernst
theorem \cite{39,40,41,42}. All unexpected experimental and theoretical
results mentioned above are valid for the boundary bodies with both
nonmagnetic \cite{26,27,28,29,30,35,36,37,38,39,40,41} and magnetic
\cite{31,32,33,34,42} metals. Many attempts have been undertaken in order
to solve this problem (see Ref.~\cite{46} for a review of different
approaches suggested in the literature).

One of this approaches addresses to the spatial nonlocality which occurs
in the screening effects or the anomalous skin effect \cite{47,48,49,50}.
The exact impedances taking the spatial nonlocality into account were
found in Refs.~\cite{48,49} for the case of nonmagnetic metals. Using
the respective reflection coefficients in the Lifshitz theory, it was
shown \cite{51,52} that the spatial nonlocality associated with the
anomalous skin effect gives only a minor contribution to the Casimir
force.

Recently the spatially nonlocal complex functions were proposed \cite{53}
which describe nearly the same response of a metal to the electromagnetic
fluctuations on the mass shell, as does the Drude model, but a significantly
different response to quantum fluctuations off the mass shell. The suggested
alternative response functions do not aim dealing with small deviations
from locality which occur for the anomalous skin effect or screening effects
\cite{47,48,49,50} in electromagnetic fields on the mass shell. They seek a
more adequate description of the quantum fluctuations off the mass shell
which are not immediately observable but contribute significantly to the
Casimir effect. The alternative response functions of Ref.~\cite{53} take the
proper account of dissipation, obey the Kramers-Kronig relations, and
describe correctly reflection of the on-shell electromagnetic
waves on metallic surfaces in optical experiments. It was shown \cite{53}
that the Lifshitz theory using the exact impedances of Refs.~\cite{48,49}
obtained from the alternative nonlocal response functions is brought
into agreement with experiments on measuring the Casimir interaction between
bodies made of nonmagnetic metal. What is more, according to the results
of Ref.~\cite{54}, the proposed alternative nonlocal response functions
bring the Lifshitz theory in agreement with the Nernst heat theorem both
for metals with perfect crystal lattices and for metals with impurities.

In this paper, a formulation of the Lifshitz theory in terms of surface
impedances, which allows an account of the spatially nonlocal dielectric
response, is extended to the case of quantization volumes bounded by
magnetic metal bodies. By solving Maxwell equations in the configuration
of an electromagnetic wave incident on a magnetic metal semispace, we
find the exact nonlocal impedances for two polarizations of the
incident field and respective reflection coefficients. The obtained
results are used to calculate the Casimir pressure between two parallel
magnetic metal (Ni) plates whose dielectric response is described by
the alternative nonlocal functions introduced in Refs.~\cite{53,54}.
It is shown that at separations of a few hundred nanometers the computed
pressures are nearly the same as are given by the Lifshitz theory using
the dissipationless plasma model. At separations of several micrometers
predictions of the Lifshitz theory using the alternative nonlocal
response are smaller in magnitude than those computed using the
plasma and Drude responses. Thus, at separation of 4 $\mu$m the Casimir
pressure computed using the alternative nonlocal response comprises
70\% and 57\% of the pressure computed using the plasma and Drude
response functions, respectively.

We have also computed the gradient of the Casimir force in the
experimental configuration of Refs.~\cite{32,33}, i.e., between a
Ni-coated sphere and a Ni-coated plate, using the alternative nonlocal
response functions at low frequencies and the available optical data
of Ni. The obtained results are shown to be in a very good agreement
with the measurement data over the entire range of separations from
223 to 550 nm. Thus, the alternative nonlocal response functions to
quantum fluctuations, which take into account the dissipation of
conduction electrons at low frequencies, bring the Lifshitz theory
in agreement with the measurement data not only for nonmagnetic metals
but for magnetic ones as well.

The paper is organized as follows. In Sec.~II, we derive the exact
impedances for magnetic media possessing the spatially nonlocal
dielectric response. Section III contains our computational results
for the Casimir pressure between two parallel magnetic metal plates
described by both nonlocal and local response functions. Section IV
presents a comparison between experiment and theory. In Sec.~V, the
reader will find our conclusions and a discussion.

\section{Exact impedances for the spatially nonlocal dielectric
response of magnetic media}

\begin{figure}[!b]
\vspace*{-3cm}
\centerline{\hspace*{-0.5cm}
\includegraphics[width=4.50in]{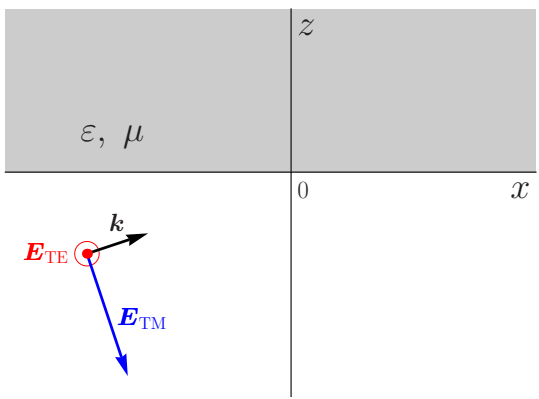}}
\vspace*{-8.cm}
\caption{\label{fg1}
Choice of the coordinate system in the configuration of
an electromagnetic wave with a wave vector $\bf {k}$ incident from
vacuum on the plane $z=0$ of magnetic medium filling in the
semispace $z>0$ (see the text for further discussion).
}
\end{figure}

We consider a magnetic metal possessing the spatially nonlocal dielectric
properties which fills in the semispace $z>0$ (see Fig.~\ref{fg1} where the
$y$ axis is directed downwards perpendicular to the $xz$ plane).
Let the wave vector $\vk=(k_x,k_y,k_z)$ of an electromagnetic wave incident
on the plane $z=0$ under some angle to the $z$-axis belongs to the $xz$ plane,
so that $k_y=0$. Then, the electric field with transverse magnetic polarization,
$\mbox{\boldmath$E$}_{\rm TM}$, is perpendicular to $\vk$ and also lies in
the $xz$ plane whereas the transverse electric field,
$\mbox{\boldmath$E$}_{\rm TE}$, is perpendicular to it and directed
downwards (see Fig.~\ref{fg1}).

The Maxwell equations inside the magnetic medium with no external charges and
currents take the standard form
\begin{eqnarray}
&&
{\rm rot}\mbox{\boldmath$E$}=-\frac{1}{c}
\frac{\partial\mbox{\boldmath$B$}}{\partial t},
\label{eq1} \\
&&
{\rm rot}\mbox{\boldmath$H$}=\frac{1}{c}
\frac{\partial\mbox{\boldmath$D$}}{\partial t},
\label{eq2} \\
&&
{\rm div}\mbox{\boldmath$B$}=0, \quad
{\rm div}\mbox{\boldmath$D$}=0,
\label{eq3}
\end{eqnarray}
\noindent
where {\boldmath$E$} is the electric field, {\boldmath$B$} is the magnetic
induction, {\boldmath$H$} is the magnetic field, and {\boldmath$D $} is the
electric displacement. With our choice of the coordinate system, all these fields
have the form
\begin{equation}
\mbox{\boldmath$F$}(t,\mbox{\boldmath$r$})=
\mbox{\boldmath$F$}(t,\mbox{\boldmath$r$};\omega,k_x)=
\mbox{\boldmath$F$}\zo e^{-i\omega t+ik_xx}.
\label{eq4}
\end{equation}
\noindent
Below we briefly repeat a derivation of the exact impedances performed in
Ref.~\cite{49} for nonmagnetic media making the corresponding generalizations
to the magnetic case where necessary.
Note that in experiments on measuring the Casimir interaction magnetic metal
is nonmagnetized in order to avoid an impact of the additional magnetic force.
In doing so our choice $k_y=0$ is not restrictive because we consider a
homogeneous isotropic medium where the preferential direction is fixed only
by the wave vector leading to tensor character of the dielectric properties
(see below). As a result, in the end of derivation one can replace $k_x$ with
$k_{\bot}=(k_x^2+k_y^2)^{1/2}$.

We start from the derivation of exact surface impedance for the TE polarization
of the electromagnetic field which is defined as \cite{48,49,52,55}
\begin{equation}
Z_{\rm TE}(\omega,\kb)=-\frac{E_y(+0;\omega,\kb)}{H_x(+0;\omega,\kb)}.
\label{eq5}
\end{equation}

For the TE-polarized field
$\mbox{\boldmath$E$}_{\rm TE}(t,\mbox{\boldmath$r$})=
\left(0,E_y(t,\mbox{\boldmath$r$}),0\right)$ and from Eq.~(\ref{eq1}) using
Eq.~(\ref{eq4}) we obtain
\begin{eqnarray}
&&
B_x\zo=\frac{ic}{\omega}\frac{dE_y\zo}{dz},
\nonumber \\
&&
B_z\zo=\frac{ck_x}{\omega}E_y\zo .
\label{eq6}
\end{eqnarray}
\noindent
{}From this it follows that both equalities in Eq.~(\ref{eq3}) are satisfied
automatically.

Now we consider the respective magnetic field
$\mbox{\boldmath$H$}(t,\mbox{\boldmath$r$})=
\big(H_x(t,\mbox{\boldmath$r$}),0,H_z(t,\mbox{\boldmath$r$})\big)$ and electric displacement
$\mbox{\boldmath$D$}_{\rm TE}(t,\mbox{\boldmath$r$})=
\big(0,D_y(t,\mbox{\boldmath$r$}),0\big)$. Using Eq.~(\ref{eq4}),
from Eq.~(\ref{eq2}) one finds
\begin{equation}
\frac{dH_x\zo}{dz}-ik_xH_z\zo=-\frac{i\omega}{c}D_y\zo .
\label{eq7}
\end{equation}

Below we assume that the effects of spatial dispersion are important for only
dielectric properties of our medium and are unrelated to its magnetic properties.
Then for the fields under consideration depending on $t$ as $\exp(-i\omega t)$
it holds
\begin{equation}
B_{x,z}\zo=\mu(\omega)H_{x,z}\zo ,
\label{eq8}
\end{equation}
\noindent
where $\mu(\omega)$ is the frequency-dependent magnetic permeability of a metal
filling the semispace $z>0$.

Substituting  Eq.~(\ref{eq8}) in Eq.~(\ref{eq7}), one obtains
\begin{equation}
\frac{1}{\mu(\omega)}\frac{dB_x\zo}{dz}-
\frac{ik_x}{\mu(\omega)}B_z\zo+\frac{i\omega}{c}D_y\zo =0.
\label{eq9}
\end{equation}
\noindent
Taking into account Eq.~(\ref{eq6}), this equation can be rewritten as
\begin{equation}
\frac{d^2E_y\zo}{dz^2}-k_x^2E_y\zo+\mu(\omega)\frac{\omega^2}{c^2}D_y\zo =0.
\label{eq10}
\end{equation}

The above equations are valid inside a medium, i.e., for $z>0$.
In order to take into account the effects of spatial dispersion, one should
use the condition of space homogeneity \cite{55,56}. To satisfy this
condition, we assume that our medium fills in not a semispace, as in
Fig.~\ref{fg1}, but all of space $-\infty<z<\infty$.
In so doing it is assumed that electrons are reflected specularly on the
plane $z=0$, i.e., the following conditions are satisfied \cite{49}:
\begin{eqnarray}
&&
E_{x,y}\zo=E_{x,y}(-z;\omega,k_x),
\nonumber \\
&&
E_{z}\zo=-E_{z}(-z;\omega,k_x),
\nonumber \\
&&
D_{x,y}\zo=D_{x,y}(-z;\omega,k_x),
\nonumber \\
&&
D_{z}\zo=-D_{z}(-z;\omega,k_x).
\label{eq11}
\end{eqnarray}

Under these conditions one can perform the Fourier transform of all fields along
the $z$-axis defined as
\begin{equation}
\widetilde{\mbox{\boldmath$F$}}\oz =\int_{-\infty}^{\infty}\!\!dz
\mbox{\boldmath$F$}\zo e^{-ik_zz}
\label{eq12}
\end{equation}
\noindent
and the inverse Fourier transform
\begin{equation}
{\mbox{\boldmath$F$}}\zo =\frac{1}{2\pi}\int_{-\infty}^{\infty}\!\!dk_z
\widetilde{\mbox{\boldmath$F$}}\oz e^{ik_zz}.
\label{eq13}
\end{equation}

Calculating the Fourier transform of both sides of Eq.~(\ref{eq10}), one obtains
\begin{equation}
I\oz-k_x^2\widetilde{E}_y\oz+\mu(\omega)\frac{\omega^2}{c^2}\widetilde{D}_y\oz=0,
\label{eq14}
\end{equation}
\noindent
where the following notation is introduced
\begin{eqnarray}
&&
I\oz\equiv \int_{-\infty}^{\infty}\!\!dz
\frac{d^2E_y\zo}{dz^2}e^{-ik_zz}
\nonumber \\
&&~~~
=\int_{0}^{\infty}\!d\left(\frac{dE_y\zo}{dz}\right)e^{-ik_zz}
\label{eq15}\\
&&~~~~~~~~~~~~~~~
+\int_{-\infty}^{0}\!d\left(\frac{dE_y\zo}{dz}\right)e^{-ik_zz}.
\nonumber
\end{eqnarray}

Integrating on the right-hand side of Eq.~(\ref{eq15}) by parts for two times with
account of Eqs.~(\ref{eq11}) and (\ref{eq12}), we find
\begin{equation}
I\oz=-k_z^2\widetilde{E}_y\oz -2\frac{dE_y(+0;\omega,k_x)}{dz},
\label{eq16}
\end{equation}
\noindent
where the last term on the right-hand side originates from a discontinuity
of the derivative $dE_y\zo/dz$ at $z=0$.

Substituting  Eq.~(\ref{eq16}) in Eq.~(\ref{eq14}), one obtains
\begin{eqnarray}
&&
-(k_x^2+k_z^2)\widetilde{E}_y\oz+\mu(\omega)\frac{\omega^2}{c^2}\widetilde{D}_y\oz
\nonumber \\
&&~~~~~~~~~~~~~~~~~~~~
=
2\frac{dE_y(+0;\omega,k_x)}{dz}.
\label{eq17}
\end{eqnarray}

On the other hand, from the first equality in Eq.~(\ref{eq6}) and Eq.~(\ref{eq8})
taken at $z=+0$ we arrive at
\begin{equation}
\frac{dE_y(+0;\omega,k_x)}{dz}=-i\mu(\omega)\frac{\omega}{c} H_x(+0;\omega,k_x).
\label{eq18}
\end{equation}

Taking into account that we deal with the TE polarization,
$\mbox{\boldmath$E$}_{\rm TE}\bot\mbox{\boldmath$k$}$, it holds \cite{55,56}
\begin{equation}
\widetilde{D}_y\oz=\Te(\omega,\vk)\widetilde{E}_y\oz,
\label{eq19}
\end{equation}
\noindent
where $\Te(\omega,\vk)$ is the transverse dielectric permittivity.

Substituting Eqs.~(\ref{eq18}) and (\ref{eq19}) in Eq.~(\ref{eq17}),
one finds
\begin{equation}
\frac{\widetilde{E}_y\oz}{H_x(+0;\omega,k_x)}=-2i
\frac{\mu(\omega){\omega}{c}}{\mu(\omega)\Te(\omega,\vk){\omega^2}-
c^2(k_x^2+k_z^2)}.
\label{eq20}
\end{equation}

For any choice of the coordinate system in the $z=0$ plane one should replace
$k_x$ with $\kb$ in Eq.~(\ref{eq20}). After this replacement, we make the
inverse Fourier transform (\ref{eq13}) on both sides of Eq.~(\ref{eq20}) and
putting $z=+0$ obtain the final result for the TE surface impedance defined
in Eq.~(\ref{eq5})
\begin{equation}
Z_{\rm TE}(\omega,\kb)=i\frac{\mu(\omega){\omega}{c}}{\pi}
\int_{-\infty}^{\infty}
\frac{dk_z}{\mu(\omega)\Te(\omega,\vk){\omega^2}-c^2\vk^2},
\label{eq21}
\end{equation}
\noindent
where $\vk^2=\skb+k_z^2$. For a nonmagnetic medium, $\mu(\omega)=1$, the
result (\ref{eq21}) coincides with that obtained in Refs.~\cite{48,49}.

We are coming now to the derivation of exact surface impedance for the TM polarization
of the electromagnetic field which is defined as \cite{48,49,52,55}
\begin{equation}
Z_{\rm TM}(\omega,\kb)=\frac{E_x(+0;\omega,\kb)}{H_y(+0;\omega,\kb)}.
\label{eq22}
\end{equation}

The TM polarized field
$\mbox{\boldmath$E$}_{\rm TM}(t,\rv)=\left(E_x(t,\rv),0,E_z(t,\rv)\right)$
has two nonzero components (see Fig.~\ref{fg1}). This makes the case of
TM polarization more complicated. Taking into account that all field components
are given by Eq.~(\ref{eq4}), one finds
$\mbox{\boldmath$B$}_{\rm TM}(t,\rv)=\left(0,B_y(t,\rv),0\right)$ and
Eq.~(\ref{eq1}) takes the form
\begin{equation}
\frac{dE_x\zo}{dz}-ik_xE_z\zo=\frac{i\omega}{c}B_y\zo .
\label{eq23}
\end{equation}

In a similar way, we have
$\mbox{\boldmath$H$}_{\rm TM}(t,\rv)=\left(0,H_y(t,\rv),0\right)$ and
$\mbox{\boldmath$D$}_{\rm TM}(t,\rv)=\left(D_x(t,\rv),0,D_z(t,\rv)\right)$
where all components are given by Eq.~(\ref{eq4}). As a result, Eq.~(\ref{eq2})
leads to
\begin{eqnarray}
&&
\frac{dH_y\zo}{dz}=\frac{i\omega}{c}D_x\zo,
\nonumber \\
&&
k_xH_y\zo =- \frac{\omega}{c}D_z\zo.
\label{eq24}
\end{eqnarray}

Taking into account that, in addition to Eq.~(\ref{eq8}), it also holds
\begin{equation}
B_y\zo=\mu(\omega)H_y\zo,
\label{eq25}
\end{equation}
\noindent
we bring Eq.~(\ref{eq24}) to the form
\begin{eqnarray}
&&
\frac{dB_y\zo}{dz}-i\mu(\omega)\frac{\omega}{c}D_x\zo=0,
\nonumber \\
&&
k_xB_y\zo +\mu(\omega)\frac{\omega}{c}D_z\zo=0.
\label{eq26}
\end{eqnarray}

We express $B_y\zo$ from the second equality in Eq.~(\ref{eq26}) and
substitute to the right-hand side of Eq.~(\ref{eq23}). The result is
\begin{equation}
ik_x\frac{dE_x\zo}{dz}+k_x^2E_z\zo-\mu(\omega)
\frac{\omega^2}{c^2}D_z\zo =0.
\label{eq27}
\end{equation}

Now we differentiate both sides of Eq.~(\ref{eq23}) with respect to $z$
and, using the first equality in Eq.~(\ref{eq26}), obtain
\begin{equation}
\frac{d^2E_x\zo}{dz^2}-ik_x\frac{\partial E_z\zo}{\partial z}+\mu(\omega)
\frac{\omega^2}{c^2}D_x\zo =0.
\label{eq28}
\end{equation}

The Fourier transform of Eq.~(\ref{eq27}) with account of Eq.~(\ref{eq11})
leads to
\begin{eqnarray}
&&
k_xk_z\widetilde{E}_x\oz-k_x^2\widetilde{E}_z\oz
\nonumber \\
&&~~~~~~~~~~~~~~~
+\mu(\omega)\frac{\omega^2}{c^2}
\widetilde{D}_z\oz=0.
\label{eq29}
\end{eqnarray}

The Fourier transform of Eq.~(\ref{eq28}) can be written in the form
\begin{equation}
I_1\oz-ik_xI_2\oz +\mu(\omega)\frac{\omega^2}{c^2}
\widetilde{D}_x\oz=0,
\label{eq30}
\end{equation}
\noindent
where the integrals
\begin{eqnarray}
&&
I_1\oz\equiv\int_{-\infty}^{\infty}\frac{d^2E_x\zo}{dz^2}
e^{-ik_zz}dz,
\nonumber \\
&&
I_2\oz\equiv\int_{-\infty}^{\infty}\frac{\partial E_z\zo}{\partial z}
e^{-ik_zz}dz
\label{eq31}
\end{eqnarray}
\noindent
are calculated similar to Eqs.~(\ref{eq15}) and (\ref{eq16}) under
conditions (\ref{eq11}) with the results
\begin{eqnarray}
&&
I_1\oz=-2\frac{dE_x(+0;\omega,k_x)}{dz}-k_z^2\widetilde{E}_x\oz ,
\nonumber \\[-1mm]
&&
\label{eq32} \\[-1mm]
&&
I_2\oz=-2E_z(+0;\omega,k_x)+ik_z\widetilde{E}_z\oz.
\nonumber
\end{eqnarray}
\noindent
The additional terms on the right-hand side of these equalities originate
from the discontinuities of the quantities $dE_x\zo/dz$ and $E_z\zo$
at $z=0$.

Substituting Eq.~(\ref{eq32}) in Eq.~(\ref{eq30}), we obtain
\begin{eqnarray}
&&
-k_z^2\widetilde{E}_x\oz+k_xk_z\widetilde{E}_z\oz
\nonumber \\
&&~~~~~~~~~
+\mu(\omega)\frac{\omega^2}{c^2}
\widetilde{D}_x\oz
\label{eq33} \\
&&~~~~~~~~~
=2\frac{dE_x(+0;\omega,k_x)}{dz}-2ik_xE_z(+0;\omega,k_x).
\nonumber
\end{eqnarray}

With account of Eq.~(\ref{eq25}), the first Maxwell equation (\ref{eq23})
taken at $z=+0$ is
\begin{equation}
\frac{dE_x(+0;\omega,k_x)}{dz}-ik_xE_z(+0;\omega,k_x)=
i\mu(\omega)\frac{\omega}{c}H_y(+0;\omega,k_x).
\label{eq34}
\end{equation}

Substituting this in Eq.~(\ref{eq33}), one obtains
\begin{eqnarray}
&&
-k_z^2\widetilde{E}_x\oz+k_xk_z\widetilde{E}_z\oz
\label{eq35} \\
&&~~
+\mu(\omega)\frac{\omega^2}{c^2}\widetilde{D}_x\oz
=2i\mu(\omega)\frac{\omega}{c}H_y(+0;\omega,k_x).
\nonumber
\end{eqnarray}

Equations (\ref{eq29}) and (\ref{eq35}) taken together give the possibility
to find the surface impedance $Z_{\rm TM}$ defined in Eq.~(\ref{eq22}).
In the presence of spatial dispersion, the quantities
$\widetilde{D}_x\oz$ and $\widetilde{D}_z\oz$ are the linear combinations of
$\widetilde{E}_x\oz$ and $\widetilde{E}_z\oz$ where the components of the
dielectric tensor serve as the coefficients \cite{55,56}
\begin{eqnarray}
&&
\widetilde{D}_x\oz=\ve_{xx}\widetilde{E}_x\oz+\ve_{xz}\widetilde{E}_z\oz,
\nonumber \\[-1.5mm]
&&
\label{eq36} \\[-1.5mm]
&&
\widetilde{D}_z\oz=\ve_{zx}\widetilde{E}_x\oz+\ve_{zz}\widetilde{E}_z\oz.
\nonumber
\end{eqnarray}

In Ref.~\cite{49} the tensor $\ve_{ij}$ was diagonalized by rotating
the coordinate system $(x,z)$ about $y$ axis by the angle $\varphi$
such that $\sin\varphi=k_x/k$, $\cos\varphi=k_z/k$.
In the rotated coordinates $(x^{\prime},z^{\prime})$ the wave vector
$\vk$ is directed along the $z^{\prime}$-axis and the dielectric tensor
takes a diagonal form with the components $\Te$ and $\Le$ where $\Le$
is the longitudinal dielectric permittivity (we omit for brevity the
arguments $\omega$ and $\vk$ in components of the dielectric tensor).

In Ref.~\cite{49} it was shown that
\begin{eqnarray}
&&
\ve_{xx}=\frac{1}{k_x^2+k_z^2}\left(\Te k_z^2+\Le k_x^2\right),
\nonumber \\
&&
\ve_{zz}=\frac{1}{k_x^2+k_z^2}\left(\Te k_x^2+\Le k_z^2\right),
\nonumber \\
&&
\ve_{xz}=\ve_{zx}=\left(\Le-\Te\right)\frac{k_xk_z}{k_x^2+k_z^2}.
\label{eq37}
\end{eqnarray}

With account of (\ref{eq36}), we rewrite Eqs.~(\ref{eq29}) and (\ref{eq35})
in the following equivalent form:
\begin{widetext}
\begin{eqnarray}
&&
\Big[k_xk_z+\mu(\omega)\frac{\omega^2}{c^2}\ve_{zx}\Big]\widetilde{E}_x\oz
+\Big[-k_x^2+\mu(\omega)\frac{\omega^2}{c^2}\ve_{zz}\Big]\widetilde{E}_z\oz=0,
\label{eq38} \\
&&
\Big[-k_z^2+\mu(\omega)\frac{\omega^2}{c^2}\ve_{xx}\Big]\widetilde{E}_x\oz
+\Big[k_xk_z+\mu(\omega)\frac{\omega^2}{c^2}\ve_{xz}\Big]\widetilde{E}_z\oz
=2i\mu(\omega)\frac{\omega}{c}H_y(+0;\omega,k_x).
\nonumber
\end{eqnarray}
\end{widetext}

By solving this system of linear equations with respect to
$\widetilde{E}_x\oz$ and using Eq.~(\ref{eq37}) for the components of a
nondiagonal dielectric tensor, we obtain
\begin{eqnarray}
&&
\frac{\widetilde{E}_x\oz}{H_y(+0;\omega,k_x)}=2i
\frac{c\omega\mu(\omega)}{k_x^2+k_z^2}
\label{eq39} \\
&&
\times
\left[\frac{k_x^2}{\mu(\omega)\omega^2\Le(\omega,\vk)} +
\frac{k_z^2}{\mu(\omega)\omega^2\Te(\omega,\vk)-c^2(k_x^2+k_z^2)}
\right].
\nonumber
\end{eqnarray}

By replacing here $k_x$ with $\kb$, as was already done in the case of the TE
polarization, and performing the inverse Fourier transform, we find the TM
surface impedance (\ref{eq22}) for a magnetic medium
\begin{eqnarray}
&&
Z_{\rm TM}(\omega,\kb)=i
\frac{c\omega\mu(\omega)}{\pi}\int_{-\infty}^{\infty}
\frac{dk_z}{\vk^2}
\left[\frac{\skb}{\mu(\omega)\omega^2\Le(\omega,\vk)}
\right.
\nonumber \\
&&~~~~~~
\left. +
\frac{k_z^2}{\mu(\omega)\omega^2\Te(\omega,\vk)-c^2\vk^2}
\right].
\label{eq40}
\end{eqnarray}
\noindent
For a nonmagnetic medium, this result coincides with respective results
of Refs.~\cite{48,49}.

For calculation of the Casimir interaction in the framework of the Lifshitz
theory (see the next section), one needs the values of surface impedances at
the pure imaginary Matsubara frequencies $i\xi_l$, where $\xi_l=2\pi k_BTl/\hbar$,
$k_B$ is the Boltzmann constant, $T$ is the temperature, and $l=0,\,1,\,2,\,\ldots$
is an integer number. Substituting $\omega=i\xi_l$ in Eqs.~(\ref{eq21}) and
(\ref{eq40}), one obtains
\begin{eqnarray}
&&
Z_{\rm TE}(i\xi_l,\kb)=\frac{c\xi_l\mu_l}{\pi}
\int_{-\infty}^{\infty}
\frac{dk_z}{\mu_l\Tle(\vk){\xi_l^2}+c^2\vk^2},
\nonumber \\
&&
Z_{\rm TM}(i\xi_l,\kb)=
\frac{c\xi_l\mu_l}{\pi}\int_{-\infty}^{\infty}
\frac{dk_z}{\vk^2}
\left[\frac{\skb}{\mu_l\xi_l^2\Lle(\vk)}
\right.
\nonumber \\
&&~~~~~~
\left. +
\frac{k_z^2}{\mu_l\xi_l^2\Tle(\vk)+c^2\vk^2}
\right],
\label{eq41}
\end{eqnarray}
\noindent
where $\Tle(\vk)\equiv\Te(i\xi_l,\vk)$,  $\Lle(\vk)\equiv\Le(i\xi_l,\vk)$,
and $\mu_l\equiv\mu(i\xi_l)$.

In terms of the surface impedances (\ref{eq41}) the amplitude reflection
coefficients on the boundary plane of magnetic metal for two polarizations
of the electromagnetic field take the form \cite{48,49,55}
\begin{eqnarray}
&&
r_{\rm TM}\xk=\frac{cq_l-\xi_lZ_{\rm TM}\xk}{cq_l+\xi_lZ_{\rm TM}\xk},
\nonumber \\
&&
r_{\rm TE}\xk=\frac{cq_lZ_{\rm TE}\xk-\xi_l}{cq_lZ_{\rm TE}\xk+\xi_l},
\label{eq42}
\end{eqnarray}
\noindent
where  $q_l\equiv{(\skb+\xi_l^2/c^2)^{1/2}}$.

Equations (\ref{eq41}) and (\ref{eq42}) make it possible to apply the Lifshitz
theory to the case of magnetic metal boundary plates possessing spatially nonlocal
dielectric response.

\section{The Casimir pressure between magnetic metal plates described by the
alternative nonlocal response functions}

As was mentioned in Sec.~I, the Lifshitz formula for the Casimir pressure $P$
between two parallel plates (semispaces) spaced at a distance $a$ was derived
within the quantum-field-theoretical and statistical approaches.
In terms of reflection coefficients on the boundary surfaces it can be written as
\cite{13,22,23}
\begin{eqnarray}
&&
P(a,T)=-\frac{k_BT}{\pi}\sum_{l=0}^{\infty}{\vphantom{\sum}}^{\prime}
\int_{0}^{\infty}\!\!\!q_l\kb d\kb
\nonumber \\
&&~~~~~~~~~~~~~~~
\times\sum_{\alpha}
\left[r_{\alpha}^{-2}\xk e^{2aq_l}-1\right]^{-1}\!\! ,
\label{eq43}
\end{eqnarray}
\noindent
where the prime on the summation sign in $l$ divides the term with $l=0$ by 2
and the sum in $\alpha$ is over two polarizations of the electromagnetic field,
$\alpha={\rm TM}$ and $\alpha={\rm TE}$.
For magnetic plates demonstrating a spatially nonlocal dielectric response the
reflection coefficients entering Eq.~(\ref{eq43}) are given by Eqs.~(\ref{eq41})
and (\ref{eq42}). Note that the Casimir pressure between metallic plates of more
than 100~nm thickness can be already considered as between semispaces and
calculated using Eq.~(\ref{eq43}) \cite{13}.

We consider the Casimir pressure between two parallel plates made of magnetic metal
Ni which is not magnetized, so that there is no magnetic force in addition to the
Casimir one. The dielectric response of Ni is supposed to be spatially nonlocal and
described by the alternative response functions introduced in Ref.~\cite{53}
\begin{eqnarray}
&&
\Te(\omega,\kb)=1-\frac{\omega_p^2}{\omega(\omega+i\gamma)}\left(
1+i\frac{\VT\kb}{\omega}\right),
\nonumber \\
&&
\Le(\omega,\kb)=1-\frac{\omega_p^2}{\omega(\omega+i\gamma)}\left(
1+i\frac{\VL\kb}{\omega}\right)^{-1} \!\!\!.
\label{eq44}
\end{eqnarray}
\noindent
Here, $\omega_p$ is the plasma frequency and $\gamma$ is the relaxation parameter
(the latter depends on $T$), and $\VT$, $\VL$ are the constants of the order of
Fermi velocity $v_F\sim 0.01c$.

The distinctive feature of response functions (\ref{eq44}) is that they nearly coincide
with the standard local Drude response function
\begin{equation}
\ve^D(\omega)=1-\frac{\omega_p^2}{\omega(\omega+i\gamma)}
\label{eq45}
\end{equation}
\noindent
for the electromagnetic fields on the mass shell. This is because
\begin{equation}
\frac{v^{\rm Tr,L}\kb}{\omega}\sim\frac{v_F}{c}\,\frac{c\kb}{\omega}
\leqslant\frac{v_F}{c}\ll 1.
\label{eq46}
\end{equation}
\noindent
As a consequence, the alternative response functions (\ref{eq44}) leads to almost
the same results, as the Drude function (\ref{eq45}), for the on-shell fields.
This is not the case, however, for the off-shell electromagnetic fields for which
the parameter (\ref{eq46}) can be large.

Although the response functions (\ref{eq44}) are of phenomenological character,
they take dissipation into account and simultaneously satisfy
the Kramers-Kronig relations and lead to an agreement of the Lifshitz theory
with experiments on measuring the Casimir interaction
between Au surfaces \cite{53}. According to the results of Ref.~\cite{54},
the Casimir entropy calculated using Eq.~(\ref{eq44}) satisfies the Nernst heat
theorem. Thus, it is of prime importance to test the alternative response functions
(\ref{eq44}) in the case of magnetic media.

For the response functions $\Tle$ and $\Lle$ depending only on $\kb$, the integrals in
Eq.~(\ref{eq41}) are easily calculated
\begin{eqnarray}
&&
Z_{\rm TE}\xk=\frac{\xi_l\mu_l}{\sqrt{c^2\skb+\mu_l\Tle(\kb)\xi_l^2}},
\label{eq47} \\
&&
Z_{\rm TM}\xk=\frac{1}{\xi_l}\left[
\vphantom{\frac{\sqrt{\mu_l\Tle(\kb)\xi_l^2}}{\Tle(\kb)}}
\frac{c\kb}{\Lle(\kb)}
\right.
\nonumber \\
&&~~~~~~~~~~~~~~~
\left.+
\frac{\sqrt{c^2\skb+\mu_l\Tle(\kb)\xi_l^2}-c\kb}{\Tle(\kb)}\right].
\nonumber
\end{eqnarray}

Substituting Eq.~(\ref{eq47})  in Eq.~(\ref{eq42}), one arrives at
\begin{widetext}
\begin{eqnarray}
&&
r_{\rm TM}\xk=\frac{q_l\Tle(\kb)-k_{\mu}^{\rm Tr}\xk-\kb[\Tle(\kb)-\Lle(\kb)]
[\Lle(\kb)]^{-1}}{q_l\Tle(\kb)+k_{\mu}^{\rm Tr}\xk+\kb[\Tle(\kb)-\Lle(\kb)]
[\Lle(\kb)]^{-1}},
\nonumber \\
&&
r_{\rm TE}\xk=
\frac{q_l\mu_l-k_{\mu}^{\rm Tr}\xk}{q_l\mu_l+k_{\mu}^{\rm Tr}\xk},
\label{eq48}
\end{eqnarray}
\end{widetext}
\noindent
where
\begin{equation}
k_{\mu}^{\rm Tr}\xk=\left[\skb+\mu_l\Tle(\kb)\frac{\xi_l^2}{c^2}\right]^{1/2}
\label{eq49}
\end{equation}
\noindent
and $\Tle$, $\Lle$ are given by Eq.~(\ref{eq44}) where one should put
$\omega=i\xi_l$.

Numerical computations of the Casimir pressure were performed by using
Eqs.~(\ref{eq43}), (\ref{eq44}), and (\ref{eq48}) at $T=300~$K.
For Ni we have used the following values of all parameters:
$\hbar\omega_p=4.89~$eV, $\hbar\gamma=0.0436~$eV \cite{57,58},
$\mu_0=110$ at $T=300~$K \cite{32,33,59}, $v_F=1.31\times 10^6~$m/s
determined in the approximation of a spherical Fermi surface, and
$\VT=\VL=7v_F$ as was used in Ref.~\cite{53} for the best agreement
between experiment and theory for Au test bodies (similar to Ref.~\cite{53}
the below results are nearly independent on the value of $\VL$ in the region
$0\leqslant\VL\leqslant 10v_F$).

It should be noted that the magnetic permeability $\mu(i\xi_l)$ quickly
decreases with $l$ and becomes equal to unity at frequencies much below the
first Matsubara frequency. Because of this, magnetic properties influence
the Casimir interaction only through the zero-frequency term of the Lifshitz
formula (\ref{eq43}) \cite{60}. In the contribution of all terms with
$l\geqslant 1$, one should put $\mu(i\xi_l)=1$. It is helpful also that
at $\xi_0=0$ the reflection coefficients (\ref{eq48}) take an especially
simple form
\begin{eqnarray}
&&
r_{\rm TM}(0,\kb)=\frac{\omega_p^2}{2\VL\gamma\kb+\omega_p^2},
\label{eq50} \\
&&
r_{\rm TE}(0,\kb)=
\frac{\mu_0\sqrt{\kb}-\sqrt{\kb+B}}{\mu_0\sqrt{\kb}+\sqrt{\kb+B}},
\nonumber
\end{eqnarray}
\noindent
where $B\equiv\mu_0\omega_p^2\VT/(\gamma c^2)$.
Interestingly, the magnetic properties make an impact only on the
TE polarization.

\begin{figure}[!b]
\vspace*{-1.5cm}
\centerline{\hspace*{-0.3cm}
\includegraphics[width=4.50in]{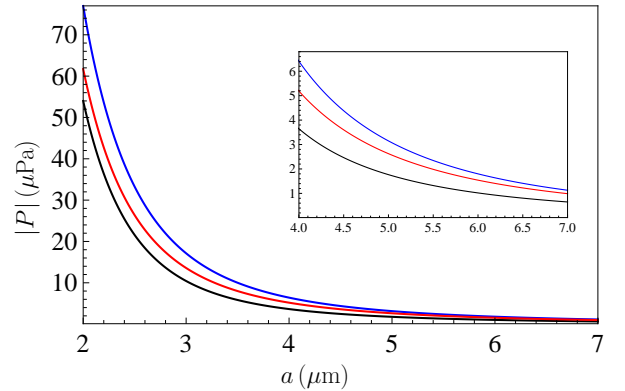}}
\vspace*{-9.5cm}
\caption{\label{fg2}
Magnitudes of the Casimir pressure between two parallel
magnetic metal plates computed using the alternative nonlocal,
plasma, and Drude response functions are shown as functions of
separation by the bottom, medium, and top lines, respectively.
The region of larger separations is shown in the inset on an
enlarged scale.
}
\end{figure}
The computational results for the magnitude of the Casimir pressure
are shown in Fig.~\ref{fg2} by the bottom line as a function of separation
in the region from 2 to $7~\mu$m. It is interesting to compare them with
similar results obtained using the standard, spatially local, response functions.
In this case we have
\begin{equation}
\Lle(\vk)=\Tle(\vk)=\ve_l=\ve(i\xi_l),
\label{eq51}
\end{equation}
\noindent
and Eq.~(\ref{eq41}) simplifies to
\begin{eqnarray}
&&
Z_{\rm TE}\xk=\frac{\xi_l\mu_l}{\sqrt{c^2\skb+\mu_l\ve_l\xi_l^2}},
\nonumber \\
&&
Z_{\rm TM}\xk=\frac{\sqrt{c^2\skb+\mu_l\ve_l\xi_l^2}}{\xi_l\ve_l}.
\label{eq52}
\end{eqnarray}

For $\mu_l=1$ these impedances were considered in Ref.~\cite{61} where it was
shown that they lead to the standard Fresnel reflection coefficients. In fact
a substitution of Eq.~(\ref{eq52}) in Eq.~(\ref{eq42}) results in
\begin{eqnarray}
&&
r_{\rm TM}\xk=\frac{q_l\ve_l-k_{\mu}\xk}{q_l\ve_l+k_{\mu}\xk},
\nonumber \\
&&
r_{\rm TE}\xk=
\frac{q_l\mu_l-k_{\mu}\xk}{q_l\mu_l+k_{\mu}\xk},
\label{eq53}
\end{eqnarray}
\noindent
where $k_{\mu}\xk$ is obtained from  $k_{\mu}^{\rm Tr}\xk$ defined  in
Eq.~(\ref{eq49}) by replacing of $\Tle$ with $\ve_l$ according to Eq.~(\ref{eq51}).
Equation (\ref{eq53}) presents the standard Fresnel coefficients commonly used in
the Lifshitz theory for both nonmagnetic ($\mu_l=1$) and magnetic plate materials.

For comparison purposes, we also compute the Casimir pressure (\ref{eq43}) between
Ni plates using the Fresnel coefficients (\ref{eq53}) and local dielectric responses
given by the dissipative Drude (\ref{eq45}) and dissipationless plasma response
functions. At the pure imaginary Matsubara frequencies these functions are given by
\begin{equation}
\ve_l^D=1+\frac{\omega_p^2}{\xi_l(\xi_l+\gamma)}, \qquad
\ve_l^p=1+\frac{\omega_p^2}{\xi_l^2}.
\label{eq54}
\end{equation}

The computational results as the functions of separation are presented in Fig.~\ref{fg2}
by the top and middle lines, respectively. In an inset, the region of larger separations
is shown on an enlarged scale. As is seen in Fig.~\ref{fg2}, the alternative nonlocal
response functions (bottom line) lead to markedly smaller theoretical values of the
pressure magnitude $|P_{nl}|$ than $|P_p|$ computed using the plasma function (middle line)
and  $|P_D|$ computed using the Drude response function over the entire range of separations
from 2 to $7~\mu$m. As an example, at $a=4~\mu$m one has $P_{nl}/P_p\approx 0.70$ and
$P_{nl}/P_D\approx0.57$. At $a=6~\mu$m the same ratios are equal to $P_{nl}/P_p\approx 0.66$
and $P_{nl}/P_D\approx0.57$.

\begin{figure}[!t]
\vspace*{-6.5cm}
\centerline{\hspace*{3cm}
\includegraphics[width=8.50in]{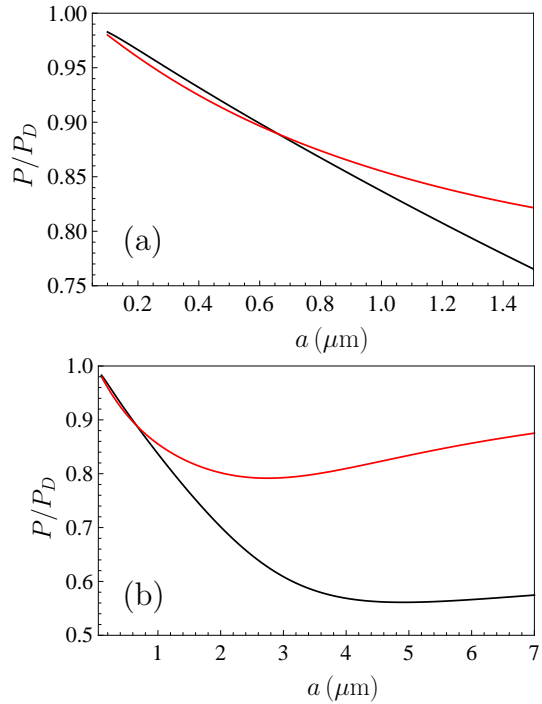}}
\vspace*{-14.5cm}
\caption{\label{fg3}
Ratios of the Casimir pressure between two parallel
magnetic metal plates computed using the alternative nonlocal
and plasma response functions to the same pressure computed
using the Drude response function ($P_{nl}/P_D$ and $P_p/P_D$,
respectively) are shown by the two lines over the separation
regions (a) from 100 nm to $1.5~\mu$m and (b) from 100 nm to
$7~\mu$m. In the region from from 100 to 655~nm the upper lines
are for $P_{nl}/P_D$ and the lower lines are for $P_p/P_D$, and
quite the reverse in the region from 655~nm to $7~\mu$m.
}
\end{figure}

In order to perform a comparison between the three response functions over a wider range
of separations, in Fig.~\ref{fg3} we plot the ratios of $P_{nl}$ and $P_p$ to $P_D$.
In so doing, we have taken into account that at separations below approximately $1~\mu$m
the response functions are influenced by the interband transitions of electrons.
An impact of these transitions becomes larger when the separation decreases.
It is included in the response functions due to conduction electrons considered above by
replacing the unities after the signs of equality
on the right-hand sides of Eqs.~(\ref{eq44}) and (\ref{eq54}) with
the appropriate function of $\xi_l$ found by means of the Kramers-Kronig relations from
the measured optical data of Ni \cite{57} (see Refs.~\cite{13,33} for details).

In Fig.~\ref{fg3}(a) the ratios $P_p/P_D$ and $P_{nl}/P_D$ are shown as functions of
separation by the lower and upper lines in the region from 100 to 655~nm, respectively.
At $a\approx 655~$nm the lines cross each other. At larger separations the ratio
$P_p/P_D$ is given by the upper line and the ratio $P_{nl}/P_D$ --- by the lower one.
In Fig.~\ref{fg3}(b) these lines are shown over the entire range of separations from
100~nm to $7~\mu$m.
Note that at separations below 100~nm theoretical predictions using all  three
response functions nearly coincide.

As is seen in Fig.~\ref{fg3}(a), within the separation region from 100 to 800~nm
the Casimir pressure between magnetic metal plates computed using the alternative
nonlocal and local plasma response functions differ by less than 1\%.
This should be compared with the fact that almost equal Casimir pressures
predicted by these response functions differ from that predicted by the Drude function
by 2\% at $a=100~$nm at by 13\% already at $a=800~$nm. According to Fig.~\ref{fg3}(b),
at separations of a few micrometers the Casimir pressures predicted by the Lifshitz
theory using all three response functions differ widely.
With further increase of separation the Casimir pressure calculated using the
alternative nonlocal and plasma response functions approach each other and the
classical limit reached in the case of plates described by the Drude function and made
of an ideal metal. This, however, holds at separations of the order of millimeters which
are immaterial due to negligibly small force values.

In the next section, we compare the theoretical predictions obtained using both local
and nonlocal response functions with the measurement data.

\section{Comparison between experiment and theory}

Experiments of Refs.~\cite{32,33} are devoted to measurements of the Casimir
interaction in the configuration of a Ni-coated hollow glass sphere with
$R=61.71~\mu$m radius and a Ni-coated Si plate. The Ni coatings
on both bodies were sufficiently thick in order they could be treated as
all-nickel when considering the Casimir interaction. These experiments were
performed in high vacuum at $T=300~$K by using the dynamic atomic force microscope
based setup operated in the frequency-shift mode. Because of this, an immediately
measured quantity was the gradient of the Casimir force between a sphere and
a plate $F_{sp}^{\prime}(a,T)=\partial  F_{sp}(a,T)/\partial a$.

According to the proximity force approximation, which is very accurate under the
condition $a\ll R$ (see below), the gradient of the Casimir force  in a sphere-plate
geometry is expressed via the Casimir pressure between two parallel plates as
\cite{11,13}
\begin{equation}
F_{sp}^{\prime}(a,T)=-2\pi RP(a,T).
\label{eq55}
\end{equation}
\noindent
This gives the possibility to compare the measurement results with theoretical
predictions of the Lifshitz theory for the Casimir pressure considered in Sec.~III.

To perform a comparison between experiment and theory, one should take into account
very small corrections to the result (\ref{eq55}) arising due to the surface roughness
on metallic coatings of a sphere and a plate \cite{11,13,62,63} and due to
deviations from the proximity force approximation \cite{64,65,66,67,68,69}.

The root-mean-square roughness on the sphere and plate surfaces was measured
using an atomic force microscope and found to be $\delta_s=1.5~$nm and
$\delta_p=1.4~$nm, respectively. So small roughness can be taken into account
perturbatively restricting ourselves to the second order in the small parameters
$\delta_{s,p}/a$. Then the theoretical force gradients (\ref{eq55}) corrected for
the presence of surface roughness are given by \cite{11,13}
\begin{equation}
F_{R}^{\prime}(a,T)=-2\pi RP(a,T)\left(1+10\frac{\delta_s^2+\delta_p^2}{a^2}
\right).
\label{eq56}
\end{equation}
\noindent
Note that at $a=300$~nm the roughness correction is equal to only 0.05\% of
the force gradient and further decreases  with increasing separation.
This is much less than the differences between alternative theoretical
predictions.

The final theoretical values of the force gradient are obtained by taking into
account the correction to the proximity force approximation
\begin{equation}
F_{\rm theor}^{\prime}(a,T)=F_{R}^{\prime}(a,T)\left[1+\theta(a,T)
\frac{a}{R}\right],
\label{eq57}
\end{equation}
\noindent
where, according to the results of Refs.~\cite{64,65,66,67,68,69}, the coefficient
$\theta(a,T)$ is negative and its magnitude does not exceed unity in the separation
region $a< 1~\mu$m. Thus, this correction is negligibly small at the experimental
separations from 225 to 550~nm. In computations below we use the same values of
$\theta(a,T)$ as in Ref.~\cite{33}.

\begin{figure*}[!t]
\vspace*{-6cm}
\centerline{
\includegraphics[width=8.50in]{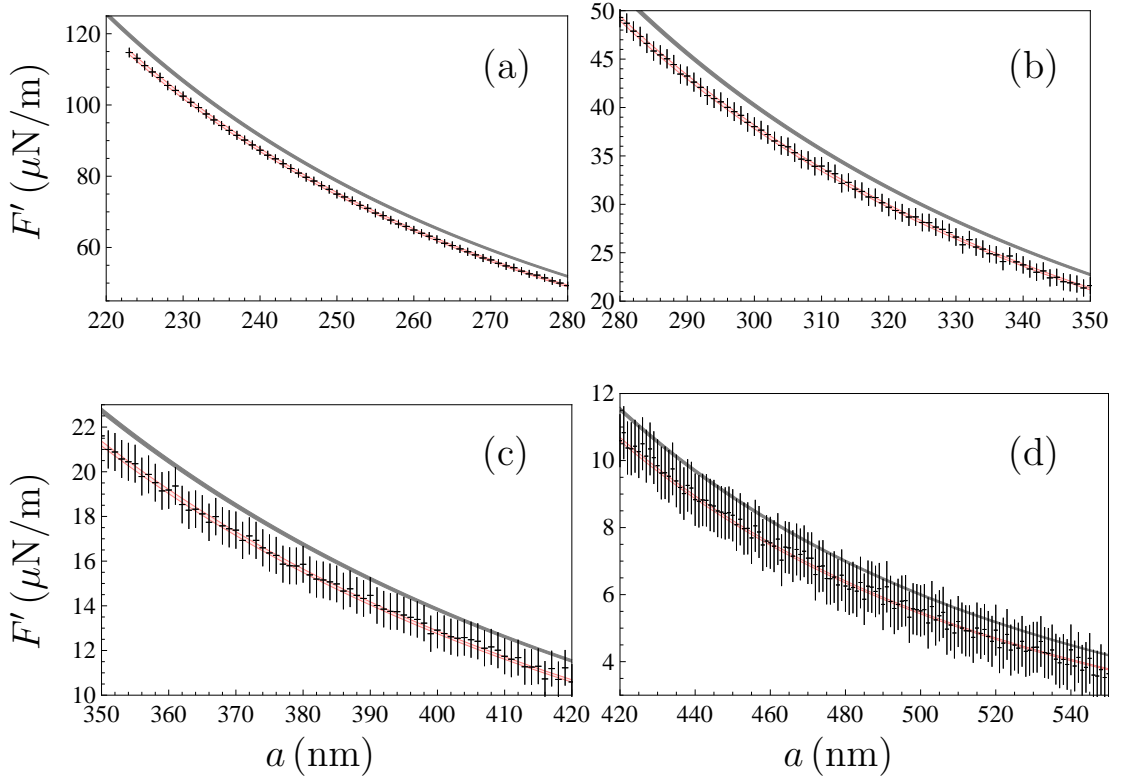}}
\vspace*{-14.5cm}
\caption{\label{fg4}
The mean measured gradients of the Casimir force between
a sphere and a plate coated with magnetic metal Ni are shown by the
crosses as functions of separation. The bottom and top theoretical
bands are computed within the Lifshitz theory using the alternative
nonlocal response functions and local Drude function, respectively.
}
\end{figure*}

Now we can compare the measurement data with theoretical predictions of the Lifshitz
theory using different response functions of magnetic metal plates.
In Figs.~\ref{fg4}(a)--\ref{fg4}(d) the mean measured data for the force gradient are shown as
crosses over the four intervals of separation distances between Ni test bodies
\cite{32}. The arms of the crosses indicate the total experimental errors determined
at a 67\% confidence level.

The theoretical predictions of the Lifshitz theory using the alternative nonlocal
response functions (\ref{eq44}, computed by Eqs.~(\ref{eq56}) and (\ref{eq57})
taking proper account of the optical data of Ni as explained in Sec.~III, are
shown in Figs.~\ref{fg4}(a)--\ref{fg4}(d) by the bottom bands. The width of these bands is
determined by the errors in all theoretical parameters, such as the plasma frequency,
relaxation parameter, sphere radius, etc. The theoretical bands computed \cite{32}
using the local dielectric response described by the plasma function $\ve_l^p$ in
Eq.~(\ref{eq54}) are indistinguishable from
the bottom ones computed using the alternative nonlocal
response functions.

As is seen in Figs.~\ref{fg4}(a)--\ref{fg4}(d), the bottom theoretical bands are in a very good
agreement with the measurement data over the entire range of separations from 223 to
550~nm. The alternative response functions, however, take into account the relaxation
properties of conduction electrons which are disregarded in an unjustified manner
when using the plasma response function.

The theoretical predictions of the Lifshitz theory computed \cite{32}
using the local dielectric response given by the Drude function $\ve_l^D$ in
Eq.~(\ref{eq54}) are shown by the top bands in Figs.~\ref{fg4}(a)--\ref{fg4}(d).
Although the Drude response function takes proper account of the relaxation
properties of conduction electrons  in the on-shell electromagnetic fields, the
theoretical predictions given by the top bands are excluded by the measurement data
over the separation region from 223 to 420~nm. This can be explained by an assumption
that the Drude function describes incorrectly the dielectric response to the
off-shell electromagnetic fields contributing to the Casimir effect.
One can conclude that the alternative nonlocal response functions provide a more
adequate response to quantum fluctuations off the mass shell.
Note that at separation distances below 100~nm the Casimir interaction is
largely caused by the contribution of interband  transitions to the dielectric
permittivity. Because of this, at so short separations the discrimination
between very close theoretical predictions obtained using the dielectric
functions $\varepsilon^D$,  $\varepsilon^p$, and $\varepsilon^{\rm Tr,L}$
is presently impossible and respective experiments are performed at larger
separations (see Fig.~\ref{fg4} and Fig.~\ref{fg5} below).

We also use another approach to a comparison between experiment and theory based
on the analysis of differences between theoretical gradients of the Casimir
force (\ref{eq57}) and mean measured gradients
\begin{equation}
\Delta F^{\prime}(a_i,T)=F_{\rm theor}^{\prime}(a_i,T)-
F_{\rm expt}^{\prime}(a_i,T),
\label{eq58}
\end{equation}
\noindent
where $a_i$ are the experimental separations at which the force gradient was measured.

\begin{figure}[!b]
\vspace*{-4.5cm}
\centerline{\hspace*{0.5cm}
\includegraphics[width=4.50in]{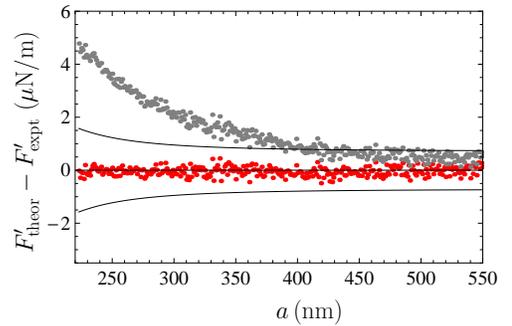}}
\vspace*{-7.8cm}
\caption{\label{fg5}
Differences between theoretical Casimir force gradients
between a sphere and a plate coated with magnetic metal Ni computed
either using the alternative nonlocal response functions (lower set
of dots) or the local Drude function (upper set of dots) and mean
experimental force gradients. The borders of the 67\% confidence
intervals for the force differences are shown by the two solid lines.
}
\end{figure}
In Fig.~\ref{fg5}, the lower set of dots presenting the quantity $\Delta F^{\prime}(a_i,T)$
as a function of separation is computed with theoretical force gradients
$F_{\rm theor}^{\prime}$ obtained using the alternative nonlocal response functions.
For the upper set of dots the gradients $F_{\rm theor}^{\prime}$  were obtained using
the local Drude response function. The two solid lines in  Fig.~\ref{fg5} indicate the
borders of the 67\% confidence intervals for the random quantity $\Delta F^{\prime}$
in Eq.~(\ref{eq58}) which take into account the total experimental and theoretical errors.

As is seen in  Fig.~\ref{fg5}, all dots belonging to the lower set are inside the confidence
intervals demonstrating a very good agreement between theory and the measurement data if
the alternative nonlocal response functions are used in computations. The same holds when
the local plasma response function is used  in computations of $F_{\rm theor}^{\prime}$
\cite{32,33} which, however, disregards the relaxation properties of conduction electrons.

{}From Fig.~\ref{fg5} it is also seen that most of dots belonging to the upper set, obtained
using the local Drude response function, are outside the confidence intervals over the
separation region from 223 to 420~nm. This means that the Lifshitz theory using the local
Drude response is experimentally excluded by measuring the Casimir interaction between magnetic
metal plates.

According to the results of Sec.~III, measurements of the Casimir interactions at separations
of a few micrometers could easily discriminate between theoretical predictions of the Lifshitz
theory obtained using the local plasma and the alternative nonlocal response functions.
This could be made, for instance, by performing the differential force measurements
proposed in Ref.~\cite{70}. At the moment, however, both these approaches to calculation
of the Casimir force are experimentally consistent and one could decide between them based on
only advantages and drawbacks in their application to a description of some other physical
phenomena.

\section{Conclusions and discussion}

In this paper, the Lifshitz theory of the Casimir force was extended
to the case of magnetic metal boundary plates possessing a spatially
nonlocal dielectric response. For this purpose, we have solved
Maxwell equations describing an electromagnetic wave incident from
vacuum on a magnetic metal semispace and expressed the exact
impedances for two independent polarizations of the electromagnetic
field via the longitudinal and transverse dielectric functions, as well
as via the magnetic permeability of a semispace metal.

The obtained results were used to calculate the Casimir pressure
between magnetic metal (Ni) plates described by the alternative
nonlocal response functions. These functions have been introduced
in Refs.~\cite{53,54} in an effort to solve puzzling problems in the
Lifshitz theory which was found to be in contradiction with the
measurement data and fundamental principles of thermodynamics when
the much studied relaxation properties of conduction electrons are
taken into account in calculations by means of the Drude response
function.

The basic idea behind introducing the alternative nonlocal response
functions is that most of the experimental information about the
electromagnetic response of a metal is obtained by using the
on-shell fields. As to a nonlocal response to the off-shell fields,
the possibilities of experimentally testing it are very limited. For
instance, some information about only the longitudinal response
function $\Le(\omega,\vk)$ can be obtained from measuring
the energy loss and momentum transfer of a beam of high energy
electrons passing through a thin metallic film \cite{55}. This doubts
on applications of the Drude response function with no modification in
the region of electromagnetic fields off the mass shell, i.e., for
$\omega^2 < k^2c^2$, which gives a sizable contribution to the
Casimir effect.

Thus, it is reasonable to look for nonlocal generalizations of the
Drude function which nearly coincide with it for the on-shell
fields but can deviate significantly for electromagnetic fluctuations
off the mass shell. Taking into account that the plasma response
function, leading to an agreement of the Lifshitz theory with
the experimental data and requirements of thermodynamics, possesses
the second order pole at zero frequency, the same property might be
expected from the sought for response. The phenomenological
alternative response functions introduced in Refs.~\cite{53,54}
satisfy these conditions.

Another motivation for using the alternative nonlocal response
functions comes from graphene. At low energies characteristic for
the Casimir effect at not too short separations, graphene is well
described by the Dirac model. In the framework of this model,
the spatially nonlocal response functions of graphene to both the
on-shell and off-shell fields can be expressed precisely based on
first principles of quantum field theory at nonzero temperature via
the components of the polarization tensor in (2+1)-dimensional
space-time (see Refs.~\cite{71,72} for the complete results).
In this situation, one expects that the Lifshitz theory of the
Casimir interaction with graphene using its exact response
functions should be in agreement with both the measurement data
and requirements of thermodynamics. These expectations
were confirmed by the measurement data of two experiments which
were found to be in excellent agreement with theoretical
predictions using the polarization tensor \cite{73,74,75,75a}. On the
other hand, the Casimir entropy in graphene systems calculated
using the polarization tensor was proven to be in perfect
agreement with the Nernst heat theorem \cite{76,77,78,79}.

After this discussion, we return to the obtained results. It was
shown that at the experimental separations from 100 to 800 nm
the Casimir pressures between two parallel Ni plates computed
by the Lifshitz formula using the alternative nonlocal and local
plasma response functions differ by less than 1\%. However, at
separations of a few micrometers these two theoretical predictions
differ between themselves and with the prediction obtained using
the local Drude function by several tens of percent. This opens
up possibilities to experimentally check these predictions in
near future.

We have also compared theoretical gradients of the Casimir
force between a Ni-coated sphere and a Ni-coated plate, computed
using the alternative nonlocal response functions and the optical
data of Ni, with the measurement data of Refs.~\cite{32,33}. The
obtained theoretical results were found in to be in a very good
agreement with the experimental ones over the entire range of
separations from 223 to 550~nm. This agreement is almost
identical to that obtained in Refs.~\cite{32,33} using the optical
data of Ni supplemented by the dissipationless plasma response
function at low frequencies \cite{32,33}. It has been known also
\cite{32,33} that the theoretical predictions obtained using the
local Drude response are excluded by the measurement data over the
range of separations from 223 to 420~nm. In so doing an advantage
of the alternative nonlocal response functions is that they take
into account the relaxation properties of conduction electrons at
low frequencies, as does the Drude function, but, as opposed to the
Drude function, leads to an agreement between experiment and theory
which could be previously reached only by using the plasma model,
i.e., by dropping the relaxation properties of conduction electrons.

In view of the above, one can conclude that the alternative nonlocal
response functions to quantum fluctuations offer certain advantages
over more conventional local response functions and deserve further
investigation.

\section*{ACKNOWLEDGMENTS}

This work was partially
supported by the Peter the Great Saint Petersburg Polytechnic
University in the framework of the Russian state assignment for basic research
(project No.\ FSEG-2020-0024).
V.~M.~M.~was partially funded by the Russian Foundation for Basic
Research, grant No.\ 19-02-00453 A.
V.~M.~M.\ was also partially supported by the Russian Government
Program of Competitive Growth of Kazan Federal University.


\end{document}